\documentclass[12pt]{article}

\usepackage[utf8]{inputenc}
\usepackage[none]{hyphenat} 
\usepackage{amsmath}
\usepackage{amssymb}
\usepackage{amsfonts}
\usepackage{multirow}
\usepackage{natbib}
\usepackage{enumerate}
\usepackage{hyperref}
\usepackage{graphicx}
\hypersetup{colorlinks=true, linkcolor=blue, urlcolor=blue, citecolor=red}
\usepackage{listings}
\usepackage{color}

\usepackage[top=2cm, bottom=2cm, left=2cm, right=2cm]{geometry}

\newtheorem{example}{Example}

\title{Enlarging of the sample to address multicollinearity}
\author{
    \bf{Román Salmerón Gómez}\thanks{Professor, Department of Quantitative methods for economics and business, University of Granada, Spain (e-mail: romansg@ugr.es).}
    \and
    \bf{Catalina García García}\thanks{Professor, Department of Quantitative methods for economics and business, University of Granada, Spain (e-mail: cbgarcia@ugr.es).}
    \and 
    \bf{Ainara Rodríguez Sánchez}\thanks{Professor, Department of Applied Economics and Economic History, National University of Distance Education, Madrid, Spain  (e-mail: arsanchez@cee.uned.es).}}
   \date{\today}

\begin{document}

\maketitle

\begin{abstract}
The paper analyzes how the enlarging of the sample affects to the mitigation of collinearity concluding that it may mitigate the consequences of collinearity related to statistical analysis but not necessarily the numerical instability. The problem that is addressed is of importance in the teaching of social sciences since it discusses one of the solutions proposed almost unanimously to solve the problem of multicollinearity.
For a better understanding and illustration of the contribution of this paper, two empirical examples are presented and not highly technical developments are used.\

keywords: \textit{enlarging the sample, multicollinearity}
\end{abstract}

\section{Introduction}

Given the model:
\begin{equation}\label{modelo1}
    \mathbf{y}=\mathbf{X}\boldsymbol{\beta}+\mathbf{u},
\end{equation}
where $\mathbf{u}$ is the random disturbance which is supposed to be spherical, $\mathbf{y}$ is the $n\times 1$ vector of the observations ot the dependent variable
and $\mathbf{X}$ is the $n\times p$ matrix ($n$ observations and $p$ variables) of the observations of the independent variables where the first column is composed of ones representing the intercept, it is well-known that ordinary least square (OLS) is commonly applied to estimate the parameter $\boldsymbol{\beta}_{n\times 1}$ of the independent variables coefficients. Among other conditions, it needs to be verified the full range condition: range of the matrix $\mathbf{X}$ equal to the number of parameters of the model or what is the same the nonexistence of perfect linear relationship between explanatory variables. If this requirement is not satisfied, the determinant of $\mathbf{X}^t\mathbf{X}$ will be zero and it will not be possible to obtain a unique solution for the estimates. This  problem is known as perfect multicollinearity. If the explanatory variables have a strong but not perfect relationship, the multicollinearity is known as near multicollinearity and in this case the estimation by OLS will be unique but the OLS estimators may present large variances, greater confidence intervals, insignificant $t$ ratios and a high coefficient of determination, as well as unstable results (\textit{small changes in the data produce wide swings in parameter estimates}, \cite{OBrien2007}), wrong signs for the estimated coefficients and difficulty in determining the individual effects of the independent variables to the dependent variable and to the coefficient of determination.

Within near multicollinearity, it is also possible to distinguish between essential collinearity and non-essential collinearity, \cite{Marquardt1980,SneeMarquardt1984}. The first one concerns the relationship between explanatory variables, excluding the intercept, while the second one regards the specific relationship between the intercept and at least one of the observed independent variables of the model. \cite{SpanosMcGuirk} reconsider the nature and consequences of near multicollinearity stating that the problem is related to two different problems that are usually mixed:
\begin{itemize}
\item The structural problem, which increases systematic volatility that can be considered a parameter problem and predictable. It is concerned with changes of the coefficient estimates associated with high correlation among the regressors, therefore it is related to the presence of high correlation among regressors. This is known as
systematic multicollinearity.
\item The numerical problem, which increases erratic volatility that is provoked by the characteristics of the data and is unpredictable.
It is concerned with the sensitivity of the coefficient estimates to proportional changes in $\mathbf{X}^t\mathbf{X}$ and $\mathbf{X}^t\mathbf{y}$, so it is related to the presence of ill-conditioning in the regressor data matrix $\mathbf{X}^t\mathbf{X}$. This is known as erratic multicollinearity
\end{itemize}

Once the degree of multicollinearity is detected, is time to consider different
solutions for its mitigation and also if it is required. \cite{Paul2006} affirms that \textit{if the goal is simply to predict $\mathbf{y}$ from a set of variables $\mathbf{X}$, then [high] multicollinearity is not a problem [because] the predictions will still be accurate, and the overall $R^2$ (or adjusted $R^2$) quantifies how well the model predicts the $\mathbf{y}$ values}.
But,even if the goal of the study were to make predictions, it is highly recommended that multicollinearity be mitigated: the researcher has to be very sure of the continuity of the relationships between explanatory variables because, if the relationship changes in the future, the forecast based on the initial model may be unreliable as well \citep{Gujarati2010,Wooldridge2008}. Thus, some strategies proposed in the literature to mitigate multicollinearity are based on create new circumstances or new models by deleting variables, introducing more information or increasing the sample or the variables used. However, it is important to take into account that these solutions can not be applied in all cases. Depending on the kind of multicollinearity some solutions could be more appropriate than others and even some of them could be inefficient or not recommended. Thus, this paper shows that the enlargement of the sample can improve the statistical analysis of the model but this improvement may not mitigate the numerical unstability of the model provoked by the multicollinearity.

This paper aims to clarify the effects of enlarging the sample as a solution for multicollinearity. The structure of the paper is as follows: Section \ref{Pre} serves as a preliminary section that reviews how to measure the degree of multicollinearity. Section \ref{example1} shows that increasing the samples may mitigate the consequences of collinearity related to statistical analysis but not necessarily the numerical instability. For this, we present an example to show that the accepted solution of increasing the sample is indicated when the sample is also improved in relation to the diversity of the data and a second example to clarify that it is possible to find situations where the increment of the sample without improving the diversity of the data may mitigate the consequences of collinearity not only in relation to the statistical analysis but also in the numerical calculation. Section \ref{conclusion} summarizes the main conclusions.

\section{How to measure the degree of multicollinearity}
\label{Pre}

As stated by \cite{Gujarati2010} or \cite{Novales1988}, the key question in an empirical analysis is not to discuss the existence of multicollinearity because it always exists, thus the dilemma is in fact the degree of multicollinearity that exists in an empirical study. Different indicators can be applied to measure the degree of multicollinearity. The variance inflation factor (VIF) is one of the most applied and can be defined as the percentage of the variance that is inflated for each coefficient:
\begin{equation}\label{vifi}
\text{VIF}_i = \frac{\widehat{Var}(\widehat{\beta}_i)}{\widehat{Var}(\widehat{\beta}_{i \mathbf{O}})} = \frac{1}{1-R^2_i},
\end{equation}
where $R^2_i$ is the coefficient of determination from regression:
\begin{equation}
    \label{reg.auxiliar}
    \mathbf{X}_i=\mathbf{X}_{-i}\boldsymbol{\alpha}+\mathbf{v},
\end{equation}
where $\mathbf{X}_i$ represents the explanatory variable under analysis and $\mathbf{X}_{-i}$ represents the matrix of all the rest of explanatory variables from the initial model. The subscript $\mathbf{O}$ represents the orthogonal (no collinearity) situation. \cite{Marquardt1970} stated that a value lower than 10 indicates no problematic collinearity, although some authors set the VIF threshold to 4 (see \cite{OBrien2007} for more details). However, the VIF presents some limitations: meanly it can not detect the non-essential multicollinearity and it can not be used for dichotomic variables (see, for example, \cite{Salmeron2020a} and \cite{Salmeron2020b}).

Alternatively, the condition number (CN) overpasses these limitations and it is defined as:
\begin{equation}\label{cni}
\text{CN}(\mathbf{X}) = \sqrt{\frac{\mu_{\text{max}}}{\mu_{\text{min}}}},
\end{equation}
where $\mu_{\text{max}}$ and $\mu_{\text{min}}$ are the higher and the lower eigenvalues respectively from matrix $\mathbf{X}^t\mathbf{X}$. Its minimum value will be one, and it appears when all the explanatory variables are orthogonal between them. Following \cite{Belsleyetal1980}, \cite{Belsley1991}, values lower than 20 light collinearity, between 20 and 30, moderate collinearity, and values higher than 30 imply strong collinearity.

\section{The enlargement of the sample and the numerical stability}
\label{example1}

One of the solutions traditionally given to multicollinearity in econometrics handbooks is the enlargement of the sample. In this sense, \cite{York}(pp.1381) presented an experiment from which stated \textit{Thus, the difference between experiments 1 and 4 clearly shows that collinearity is not necessarily a problem if we still have a sufficient absolute amount of information to estimate the independent effects of a given variable}. However, we cannot agree with this statement.

Our first consideration is that the contribution of \cite{York} is illustrated with a hypothetical experimental design where explanatory variables are defined as binary variables and with these kinds of variables it makes no sense to calculate the coefficient of correlation since it is only applicable to quantitative variables. Thus, there is also no utility in calculating the variance inflation factor (VIF). For this reason, we recommend using other measures to determine if the degree of collinearity in the model is worrisome, such as the condition number (CN). Appendix \ref{CN.York} shows that experiments 2, 3 and 4 of York's example 1 present the same value for the CN, concluding that the degree of collinearity is light.

Second, \cite{OBrien} showed that an increase in the size of the sample can mitigate the consequences of approximate collinearity due to the decrease in the variance of the estimated coefficient.
This author states that \textit{more the same can certainly be hepful; a new analysis based on the additional cases of the same sort should provide more reliable point estimates and smaller confidence intervals}.
However, it may not mitigate the instability that collinearity provokes on the estimations of the coefficients of explanatory variables since \textit{this sensitivity is due to collinearity, not sample size}, see \cite{WinshipWestern}.
Thus, consequences of collinearity may not be mitigated if the increment of the sample is obtained only adding the same information of the initial sample.
Indeed, \cite{WinshipWestern} noticed that \textit{multicollinearity may be particu\-lar\-ly dangerous with large samples sizes}.

It is true that neither \cite{York} nor \cite{OBrien} refer to the problem of numerical stability, either because they assume that it does not appear or they do not consider it relevant. However, we consider that not mentioning this possibility may lead to misinterpretations in researchers unfamiliar with the multicollinearity problem, so it is relevant to analyze this possibility.
The following example reproduces the idea of York's example 1, incrementing the size of the sample and repeating exactly the same data a determined number of times
analyzing the effect of the enlargement of the sample on the statistical analysis  (in terms of the estimated variance of the regression coefficients) and on the numerical analysis (in terms of unstable results) of the model (\ref{modelo1}).
Both examples show that although the enlargement of the sample will reduce the variance of the regressions coefficients and, consequently, the experimental value of the individual significance test increases, it is not recommendable to establish conclusions from the obtained results until it is verified that there is no numerical instability in the model.

\begin{example}
\label{example.Wissell}
In Table \ref{tab3}, Model 1 shows the estimation by ordinary least squares (OLS) that analyzes the effect of consumption ($\mathbf{C}$) and personal incomes ($\mathbf{I}$) on outstanding mortgage debt ($\mathbf{D}$) from 1996 to 2012. Note that none of the estimated coefficients are significantly different from zero, although the model is globally significant. This conclusion is contradictory and a symptom of the possible existence of worrisome collinearity in the model.

\begin{table}
\centering
\begin{tabular}{cccc}
\hline
Variable & Model 1 & Model 2 & Model 3 \\
\hline
Intercept & -0.1174 & -0.1174 & -4.74586***\\
& (6.4764)& ( 2.1012)&(1.62326)\\
Consumption & -2.3429 & -2.3429* & 0.02951\\
&(3.33507)&(1.0820)& (0.81048)\\
Personal incomes & 2.8562 & 2.8562** &1.49682***\\
&(1.91234)&(0.6204)& (0.4625)\\
\hline
$R^{2}$ & 0.92202 & 0.92202&0.9158\\
$\widehat{\sigma}^{2}$ & 0.8228 &0.6928898&0.748571\\
$F_{2,14}$ & 82.77&786.3&723 \\
\hline
\end{tabular}
\caption{OLS estimation for the data set employed in \cite{Wissell} (Model 1). The OLS estimation is performed on the initial sample repeated 7 times (N=136) (Model 2) and on the initial sample repeated 7 times (N=136) and perturbed by 1\% (Model 3). (*** indicates 99\% confidence level, ** indicates 95\% confidence level and * indicates 90\% confidence level.)} \label{tab3}
\end{table}

Since collinearity is a problem inherent to the sample, the first solution would be to enlarge the sample and/or to improve its quality. However, it is possible that these measures were not taken in the beginning since it was either not possible to acquire them or they entailed a high material or economic cost. Nevertheless, we have considered that it is possible to enlarge the sample. If the sample was initially excessively biased (as it is in the case of the erratic collinearity in this example), then the enlargement of the sample could mitigate the collinearity. However, if the enlargement of the sample is obtained with information similar to the initial sample, the collinearity could not be mitigated, although the prediction could be improved. An extreme case would be to enlarge the sample only repeating exactly the same information of the initial sample. Thus, we enlarged the sample of \cite{Wissell} (Model 1) by repeating the same information 7 times (N=136). The results are shown in Model 2 in Table \ref{tab3}.

Note that the estimations of the coefficients of the independent variables and the coefficients of determination are the same, while the standard deviation of the estimated coefficients has decreased and the experimental statistic of the global significance has increased. Appendix \ref{teoria.aumentar.muestra} shows that this result will always be obtained when the sample is enlarged in this way.

What has happened? Observing the following expression of the variance of the $j-$estimated coefficient\footnote{Note that $\sigma^{2}$ is the variance of the random disturbance, and $R^{2}_{j}$ is the coefficient of determination of the auxiliary regression with variable $-j$ as the dependent variable, $\mathbf{X}_{j}$, as a function of the rest of the independent variables of the initial model}:
$$var \left( \widehat{\boldsymbol{\beta}}_{j} \right) = \frac{\sigma^{2}}{n \cdot var (\mathbf{X}_{j}) \cdot (1 - R^{2}_{j})},$$
where it is evident that an increase in the sample implies a decrease in the estimated standard deviation which implies an increment in the experimental statistic of individual significance. Thus, the tendency to reject the null hypothesis of nullity is reversed. Therefore, the initial contradiction found in this example would have been solved.

However, this model, which is estimated from the augmented samples, presents a VIF equal to 262.858 and a CN equal to 207.6262. Thus, both measures maintain the values of the model with the initial sample (see Appendix \ref{teoria.aumentar.muestra}). These values indicate the existence of collinearity, and seem to be a contradiction of the improvement shown in relation to the individual significance.

As has been commented, it is possible to distinguish two problems related to approximate collinearity: one referring to the statistical analysis of the estimated coefficients and the other referring to the numerical calculation of the same. With the enlargement of the sample, we have only mitigated the statistical analysis.
To analyze the numerical stability, we will study whether small changes in the data lead to large variations in the estimates. With such goal, we have included a perturbation of 1\% in the independent variables consumption and personal incomes with the same procedure shown by \cite{Belsley1984}. Thus, given a vector $\mathbf{x}$ of size $n$, a perturbation of 1\% is obtained by the following expression:
$$\mathbf{x}_{p} = \mathbf{x} + 0.01 \cdot \mathbf{p} \cdot \frac{||\mathbf{x}||}{||\mathbf{p}||},$$
where $\mathbf{p}$ is a random vector with the same dimension as $\mathbf{x}$ and $||\mathbf{x}|| = \sqrt{\sum \limits_{i=1}^{n} x_{i}^{2}}$.

Table \ref{tab3} shows Model 3, in which the OLS estimations of the model are performed on the sample that is enlarged and perturbed. Note that the inclusion of a perturbation of 1\% in the independent variables implies a change in the estimations of\footnote{The subindex $p$ refers to the estimation of the perturbed model.} $\frac{||\boldsymbol{\beta} - \boldsymbol{\beta}_{p}||}{||\boldsymbol{\beta}||} = 145.4431\%$.

In addition, this numerical change is accompanied by statistical changes since the coefficient of the intercept becomes significantly different than zero and the coefficient of the consumption does not become significantly different from zero.

Finally, note that if the process of perturbing and estimating the model is repeated 1000 times, an average variation of the estimated coefficient of the independent variables is obtained that is equal to 78.25003\% with a standard deviation equal to 31.03021\% (minimum and maximum values equal to 1.022916\% and 174.0052\%, respectively). That is, the instability of the estimations of the model with the augmented sample is not a consequence of a concrete case.
\hfill $\Box$
\end{example}

The previous example illustrates that the widely accepted solution of increasing the sample is indicated when the increment implies an improvement of the sample in relation to the variability and the diversity of the data. Otherwise, only the consequences of collinearity related to the inference of the model will be mitigated and not the consequences related to the numerical calculation. Thus, by following the classification presented by \cite{SpanosMcGuirk}, incrementing the sample can mitigate the erratic collinearity but it may not affect the systematic collinearity. Thus, even in this case, if the estimations are unstable, it would not be recommended to obtain conclusions from the individual significance test, although the standard deviation of the estimated coefficient has decreased since \textit{the standard errors for these regression estimates can be made arbitrarily small by choosing $n$ to be sufficiently large}, see \cite{WinshipWestern}. For example, the results presented in Model 2 and Model 3 (Table \ref{tab3}) lead to different conclusions.

To conclude, it is necessary to clarify that it is possible to find situations where the increment of the sample, even repeating the same information of the initial sample, may mitigate the consequences of collinearity not only in relation to the statistical analysis but also in the numerical calculation. This possibility is illustrated in the following example.

\begin{example}
\label{example.KG}
From the data set employed by \cite{KleinGoldberger}, Table \ref{tab3.bis} presents in Model 1 the influence of the wage incomes ($\mathbf{I}$), non-farm incomes ($\mathbf{InA}$) and farm incomes ($\mathbf{IA}$) on consumption ($\mathbf{C}$) from 1936 to 1952 (years 1942, 1943 and 1944 are missing due to the war).

\begin{table}
\centering
\begin{tabular}{cccc}
\hline
Variable & Model 1 & Model 2 & Model 3\\
\hline
Intercept& 18.7021*** & 18.7021*** & 18.0818***\\
& (6.84544)& (1.27115)&(1.2861)\\
Wage income & 0.3803 &0.3803***& 0.3761*** \\
& (0.3121)& (0.05796)&(0.0569)\\
Non-farm incomes & 1.4186 & 1.4186***& 1.4294*** \\
& (0.7204)& (0.13377)&(0.1312)\\
Farm incomes & 0.5331 & 0.5331*& 0.5514* \\
& (1.399)& (0.25994)&(0.2614)\\
\hline
$R^{2}$ & 0.9187&0.9187&0.9183 \\
$\widehat{\sigma}^{2}$ & 36.7236&26.59465&26.7289 \\
$F_{3,10}$ & 37.68&1093&1086 \\
\hline
\end{tabular}
\caption{OLS estimation for the data set employed in \cite{KleinGoldberger} (Model 1), where the OLS estimation is performed on the initial sample repeated 20 times (N=294) (Model 2) and on the initial sample repeated 20 times (N=294) and perturbed by 1\% (Model 3). (*** indicates 99\% confidence level, ** indicates 95\% confidence level and * indicates 90\% confidence level.)} \label{tab3.bis}
\end{table}

Note that none of the estimated coefficients are significantly different than zero although the model is globally significant, which is contradictory. The determinant of the correlation matrix presents a value equal to 0.03713592, the VIFs are equal to 12.296544, 9.230073 and 2.976638, and the CN is equal to 35.88644. With this information, it is possible to conclude that there is worrisome collinearity in the model.

Table \ref{tab3.bis} shows in Model 2 the estimation of the model from the enlarged sample that is repeated 20 times using the initial information. Note that all the estimated coefficients are now significantly different than zero due to a relevant decrease of the estimated standard deviation.

The question is whether the enlargement of the sample has mitigated not only the consequence on the statistical analysis but also the consequences on the numerical calculation. To answer this question, we introduce a perturbation of 1\% in the independent variables and estimate by OLS, obtaining the results displayed in Model 3 (Table \ref{tab3.bis}). In this case, the estimations of the coefficients are very similar to those shown in Model 2 (Table \ref{tab3.bis}). Indeed, there is only a variation of 3.306791\%.

Finally, if the process of perturbation and estimation is repeated 1000 times, the average variation in the estimation of the coefficients of the independent variables is equal to 2.497387\%, with a standard deviation equal to 0.5000712\% (minimum and maximum values are equal to 0.7868816\% and 4.187891\%, respectively). That is, the stability of the estimations of the model with the augmented sample is not a consequence of a concrete case.
\hfill $\Box$
\end{example}

\section{Conclusions}
\label{conclusion}

The consequences of worrisome collinearity in a linear regression model can be related to the numerical instability in the estimations of the coefficients of the independent variables or related to the inflated variance of those estimators and its repercussion in the statistical analysis of the model, mainly the individual inference.

Section \ref{example1} shows that enlarging the size of the sample may mitigate the consequences related to the statistical analysis (inference) but not necessarily the numerical instability, especially if the collinearity is systematic. Note that the conclusion should be based on firm statistical results and not subject to statistical instability when the data are subject to serious collinearity. Thus, the problem of multicollinearity is being separated from a problem of small sample size (contrarily to the paper presented by \cite{Goldberger}). At the same time, it is noted that multicollinearity may be a problem even in those cases with a large sample size.

We recommend introducing a perturbation in the original model and in the model where the collinearity has supposedly been mitigated. With this procedure, it is possible to check if the variation in the estimations of the coefficients is high or not and consequently quantify the numerical instability. An interesting research line could be to determine a threshold from which to conclude that the variation provoked by the perturbation is worrisome.

\bibliographystyle{chicago}
\bibliography{bib}

\appendix
\section*{APPENDIX}

\section{Condition number for York's example 1}
\label{CN.York}

Given the multiple linear model (\ref{modelo1}), experiments 2, 3 and 4 of York's example 1 present the following matrix of design:
$$\mathbf{X}_{n \times 3} = \left(
\begin{array}{ccc}
\mathbf{1}_{49 \cdot m \times 1} & \mathbf{0}_{49 \cdot m \times 1} & \mathbf{0}_{49 \cdot m \times 1} \\
\mathbf{1}_{m \times 1} & \mathbf{1}_{m \times 1} & \mathbf{0}_{m \times 1} \\
\mathbf{1}_{m \times 1} & \mathbf{0}_{m \times 1} & \mathbf{1}_{m \times 1} \\
\mathbf{1}_{49 \cdot m \times 1} & \mathbf{1}_{49 \cdot m \times 1} & \mathbf{1}_{49 \cdot m \times 1}
\end{array} \right),$$
where $\mathbf{0}$ and $\mathbf{1}$ are, respectively, a vector or appropriate dimensions with zeros and one, and $n = 100 \cdot m$. Experiment 2 is obtained for $m=1$ , experiment 3 for $m=10$, and experiment 4 for $m=100$.

Transforming this matrix to be unit length, it is obtained that:
$$\widetilde{\mathbf{X}}_{n \times 3} = \left(
\begin{array}{ccc}
\frac{\mathbf{1}_{49 \cdot m \times 1}}{\sqrt{100 \cdot m}} & \frac{\mathbf{0}_{49 \cdot m \times 1}}{\sqrt{50 \cdot m}} & \frac{\mathbf{0}_{49 \cdot m \times 1}}{\sqrt{50 \cdot m}} \\
\frac{\mathbf{1}_{m \times 1}}{\sqrt{100 \cdot m}} & \frac{\mathbf{1}_{m \times 1}}{\sqrt{50 \cdot m}} & \frac{\mathbf{0}_{m \times 1}}{\sqrt{50 \cdot m}} \\
\frac{\mathbf{1}_{m \times 1}}{\sqrt{100 \cdot m}} & \frac{\mathbf{0}_{m \times 1}}{\sqrt{50 \cdot m}} & \frac{\mathbf{1}_{m \times 1}}{\sqrt{50 \cdot m}} \\
\frac{\mathbf{1}_{49 \cdot m \times 1}}{\sqrt{100 \cdot m}} & \frac{\mathbf{1}_{49 \cdot m \times 1}}{\sqrt{50 \cdot m}} & \frac{\mathbf{1}_{49 \cdot m \times 1}}{\sqrt{50 \cdot m}}
\end{array} \right),$$
and, then, the matrix:
$$\widetilde{\mathbf{X}}^{t} \widetilde{\mathbf{X}} = \left(
\begin{array}{ccc}
1 & \frac{50 \cdot m}{\sqrt{100 \cdot m \cdot 50 \cdot m}} & \frac{50 \cdot m}{\sqrt{100 \cdot m \cdot 50 \cdot m}} \\
\frac{50 \cdot m}{\sqrt{100 \cdot m \cdot 50 \cdot m}} & 1 & \frac{49 \cdot m}{50 \cdot m} \\
\frac{50 \cdot m}{\sqrt{100 \cdot m \cdot 50 \cdot m}} & \frac{49 \cdot m}{50 \cdot m} & 1
\end{array} \right) =
\left(
\begin{array}{ccc}
1 & \sqrt{0.5} & \sqrt{0.5} \\
\sqrt{0.5} & 1 & 0.98 \\
\sqrt{0.5} & 0.98 & 1
\end{array} \right),$$
where the following eigenvalues are presented: 2.6035978, 0.3764022 and 0.02. Thus, the condition number is the same for the three experiments and equal to $\sqrt{\frac{2.6035978}{0.02}} = 11.40964.$
Thus, the degree of collinearity existing in these experiments is light.

\section{Enlarging of the sample size in the multiple linear regression}
\label{teoria.aumentar.muestra}
Given the observations, $(\mathbf{y}, \mathbf{X})$, of the variables of model (\ref{modelo1}) where a worrisome degree of collinearity is supposed to exist, the following enlarged model is proposed:
\begin{equation}
\label{model0.aumentado}
\mathbf{y}_{A} = \mathbf{X}_{A} \cdot \boldsymbol{\beta} + \mathbf{v},
\end{equation}
where $\mathbf{v}$ is the random disturbance (supposedly spherical) and:
$$\mathbf{y}_{A} = \left(
\begin{array}{c}
\mathbf{y} \\
\vdots \\
\mathbf{y}
\end{array} \right)_{n \cdot h \times 1}, \quad
\mathbf{X}_{A} = \left(
\begin{array}{c}
\mathbf{X} \\
\vdots \\
\mathbf{X}
\end{array} \right)_{n \cdot h \times k},$$
that is, $(\mathbf{y}_{A}, \mathbf{X}_{A})$ is obtained by repeating $h$ times the initial data $(\mathbf{y}, \mathbf{X})$.

It is easy to show that the estimator of $\boldsymbol{\beta}$ in model (\ref{model0.aumentado}), denoted as\footnote{In general, the subindex $A$ is used to refer to model (\ref{model0.aumentado}), and without the subindex to refer to model (\ref{modelo1}).} $\widehat{\boldsymbol{\beta}}_{A}$, verifies the following characteristics:
\begin{eqnarray}
\widehat{\boldsymbol{\beta}}_{A} &=& \left( \mathbf{X}_{A}^{t} \mathbf{X}_{A} \right)^{-1} \mathbf{X}_{A}^{t} \mathbf{y}_{A} = \frac{1}{h} \cdot \left( \mathbf{X}^{t} \mathbf{X} \right)^{-1} \cdot h \cdot \mathbf{X}^{t} \mathbf{y} = \widehat{\boldsymbol{\beta}}, \nonumber \\
SCR_{A} &=& \mathbf{y}_{A}^{t} \mathbf{y}_{A} - \widehat{\boldsymbol{\beta}}_{A}^{t} \mathbf{X}_{A}^{t} \mathbf{y}_{A} = h \cdot \mathbf{y}^{t} \mathbf{y} - \widehat{\boldsymbol{\beta}}^{t} \cdot h \cdot \mathbf{X}^{t} \mathbf{y} = h \cdot SCR, \nonumber \\
SCT_{A} &=& \mathbf{y}_{A}^{t} \mathbf{y}_{A} - n \cdot h \cdot \overline{\mathbf{y}}^{2}_{A} = h \cdot \mathbf{y}^{t} \mathbf{y} - n \cdot h \cdot \overline{\mathbf{y}}^{2} = h \cdot SCT, \nonumber \\
R^{2}_{A} &=& 1 - \frac{SCR_{A}}{SCT_{A}} = R^{2}, \label{R2.aumentado} \\
\overline{R}_{A}^{2} &=& 1 - (1 - R^{2}_{A}) \cdot \frac{n \cdot h - 1}{n \cdot h - k} = 1 - (1 - \overline{R}^{2}) \cdot \frac{n-k}{n-1} \cdot \frac{n \cdot h - 1}{n \cdot h - k}, \nonumber \\
\widehat{\sigma}^{2}_{A} &=& \frac{SCR_{A}}{n \cdot h - k} = h \cdot \frac{n-k}{n \cdot h - k} \cdot \widehat{\sigma}^{2}, \nonumber \\
\widehat{var \left( \widehat{\boldsymbol{\beta}}_{A} \right)} &=& \widehat{\sigma}^{2}_{A} \cdot \left( \mathbf{X}_{A}^{t} \mathbf{X}_{A} \right)^{-1} = \frac{n-k}{n \cdot h - k} \cdot \widehat{var \left( \widehat{\boldsymbol{\beta}} \right)}, \label{var.aumentada} \\
t_{exp,A} \left( \beta_{i} \right) &=& \left| \frac{\widehat{\beta}_{i,A}}{\sqrt{\widehat{var (\widehat{\beta}_{i,A})}}} \right| = \left| \frac{\widehat{\beta}_{i}}{\sqrt{\frac{n-k}{n \cdot h - k}} \cdot \sqrt{\widehat{var (\widehat{\beta}_{i})}}} \right| = \sqrt{\frac{n \cdot h - k}{n - k}} \cdot t_{exp} \left( \beta_{i} \right), \label{texp.aumentada} \\
F_{exp,A} &=& \frac{\frac{R^{2}_{A}}{k-1}}{\frac{1-R^{2}_{A}}{n \cdot h - k}} = \frac{n \cdot h - k}{k-1} \cdot \frac{R^{2}}{1 - R^{2}} = \frac{n \cdot h - k}{n-k} \cdot F_{exp}. \label{Fexp.aumentada}
\end{eqnarray}

Since $h>1$, then $\frac{n\cdot h - k}{n-k} > 1$, and from expressions (\ref{var.aumentada}), (\ref{texp.aumentada}) and (\ref{Fexp.aumentada}), it is obtained that:
$$\widehat{var \left( \widehat{\boldsymbol{\beta}}_{A} \right)} < \widehat{var \left( \widehat{\boldsymbol{\beta}} \right)}, \quad
t_{exp,A} \left( \beta_{i} \right) > t_{exp} \left( \beta_{i} \right), \quad
F_{exp,A} > F_{exp},$$
that is, the estimated variance of the estimators is reduced in model (\ref{model0.aumentado}) compared to (\ref{modelo1}), while the individual and global experimental statistics are increased.

To reject the null hypothesis in the individual significance test in model (\ref{model0.aumentado}), it has to be verified that\footnote{Note that in comparison with model (\ref{modelo1}), the theoretical statistics are also reduced.}:
$$t_{exp,A} \left( \beta_{i} \right) > t_{n \cdot h - k} (1 - \alpha/2),$$
which is equivalent to verifying the condition $h_{i} > bound_{i}$ where:
\begin{equation}
\label{cota.aumentar}
bound_{i} = \frac{1}{n} \left( \left( \frac{t_{n \cdot h - k} (1 - \alpha/2)}{t_{exp} \left( \beta_{i} \right)} \right)^{2} \cdot (n-k) + k \right).
\end{equation}
Thus, it will reject the null hypothesis that the coefficient $\beta_i$ is null if the sample is increased $h_{i}$ times.

Observing expression (\ref{cota.aumentar}), note that to calculate $bound_{i}$, it is necessary to know $h$ to determine the value of $t_{n \cdot h - k} (1 - \alpha/2)$. This contradiction is resolved by approximating the value of $t_{n \cdot h - k} (1 - \alpha/2)$ as 1.96 using the relation that exists between the typical normal and the Student’s $t$ distributi\-ons.

Consequently, considering $h = max \{ h_{1}, \dots, h_{k} \}$, it is assured that the null hypothesis is rejected in all individual significance tests of model (\ref{model0.aumentado}). In example \ref{example.Wissell}, without considering the intercept, it is calculated that $h=7$, and in the example \ref{example.KG}, it is obtained that $h=20$.

Thus, it is clear that one of the symptoms of collinearity is mitigated: the tendency of not rejecting the null hypothesis of the individual significance and rejecting the null hypothesis of the global significance test.

However, if the most common measures to diagnose collinearity are calculated in model (\ref{model0.aumentado}), it is obtained that:
\begin{itemize}
\item The calculation of VIF is based on the coefficient of determination of the auxiliary regression of variable $j$ of matrix $\mathbf{X}_{A}$ as a function of the rest of the variables of this same matrix. However, this auxiliary regression is the augmented version of the auxiliary regression used to calculate the VIF in model (\ref{modelo1}). As exposed in (\ref{R2.aumentado}), the coefficients of both auxiliary regressions coincide and, consequently, the VIF of the initial and augmented model will also be the same.
\item Due to $\mathbf{X}_{A}^{t} \mathbf{X}_{A} = h \cdot \mathbf{X}^{t} \mathbf{X}$, the eigenvalues of $\mathbf{X}_{A}^{t} \mathbf{X}_{A}$ and $\mathbf{X}^{t} \mathbf{X}$ are proportional, that is, if $\mu$ is an eigenvalue of $\mathbf{X}^{t} \mathbf{X}$, $h \cdot \mu$ is an eigenvalue of $\mathbf{X}_{A}^{t} \mathbf{X}_{A}$. In this case:
$$CN(\mathbf{X}_{A}) = \sqrt{\frac{h \cdot \mu_{max}}{h \cdot \mu_{min}}} = CN(\mathbf{X}),$$
that is, the CN in both models will coincide (this fact was also exposed in appendix \ref{CN.York}). Note that this relation is also verified when transforming the matrix $\mathbf{X}_{A}$ to be unit length.
\end{itemize}
Thus, if these measures indicate initially that there is collinearity in model (\ref{modelo1}), the conclusion will be the same in the augmented model (\ref{model0.aumentado}). Then, although one of the symptoms of collinearity may be mitigated, the diagnostic measure will continue indicating the existence of collinearity.
That is, \textit{we can have a sample of infinity size, sampling standard errors of zero, and we would have the same problem. This suggests that multicollinearity may actually be most dangerous with large data sets} (\cite{WinshipWestern}).

\end{document}